\documentclass[aps,amsmath,amssymb,prl,twocolumn,superscriptaddress]{revtex4}

\usepackage{amsfonts}
\usepackage{graphicx}

\begin{document}

\title{Zero temperature Dephasing and the Friedel Sum Rule}

\author{Bernd Rosenow}
\affiliation{Institut f\"ur Theoretische Physik, Universit\"at Leipzig, D-04103, Leipzig, Germany}
\author{Yuval Gefen}
\affiliation{Department of Condensed Matter Physics, The Weizmann Institute of Science,  Rehovot 76100, Israel.}

\date{November 24, 2011}

\begin{abstract}
Detecting  the passage of an interfering particle through one of the 
interferometer's arms, known as ``which path'' measurement, gives rise to interference visibility  degradation (dephasing). Here  we consider a detector at {\em equilibrium}. At finite temperature dephasing is caused by thermal fluctuations of the detector. More interestingly,  in the zero temperature limit, equilibrium  quantum fluctuations of the detector  give rise to dephasing of the out-of-equilibrium interferometer.  This dephasing is a manifestation of an orthogonality catastrophe which differs qualitatively from Anderson's. Its magnitude is directly related to the Friedel sum rule.

\end{abstract}

\maketitle

Electronic interferometry has been invaluable in studying fundamental quantum mechanical phenomena such as the Aharonov-Bohm effect, two-particle Hanbury-Brown and Twiss correlations, and fractional statistics  of Abelian and non-Abelian anyons. The visibility of the interferometer's signal is suppressed when  the trajectory of the interfering particle is measured \cite{Stern+90}. This  is akin to ``dephasing'' of the system at hand. The physics of such ``which path'' measurement has been observed experimentally \cite{whichpath}
and discussed theoretically \cite{Gurvitz97, Levinson97,Aleiner+97}. Standard detection schemes involve an  out-of-equilibrium detector, e.g.~a current-carrying quantum point contact, electrostatically coupled to the interferometer. The dephasing is directly related to entanglement between  the state of the system and that of the detector 
\cite{Neder+07}.

What happens when the detector is set to be at equilibrium? It is known that coupling to a dissipative environment may give rise to dephasing through  thermal fluctuations \cite{AlArKh81,MaBr02,SeBu01,Delft04} and to  effective  mass renormalization \cite{HoDo10}. But can coupling to a detector lead to dephasing at zero temperature?

\begin{figure}[tbh]
\centering\includegraphics[width=0.9\columnwidth]{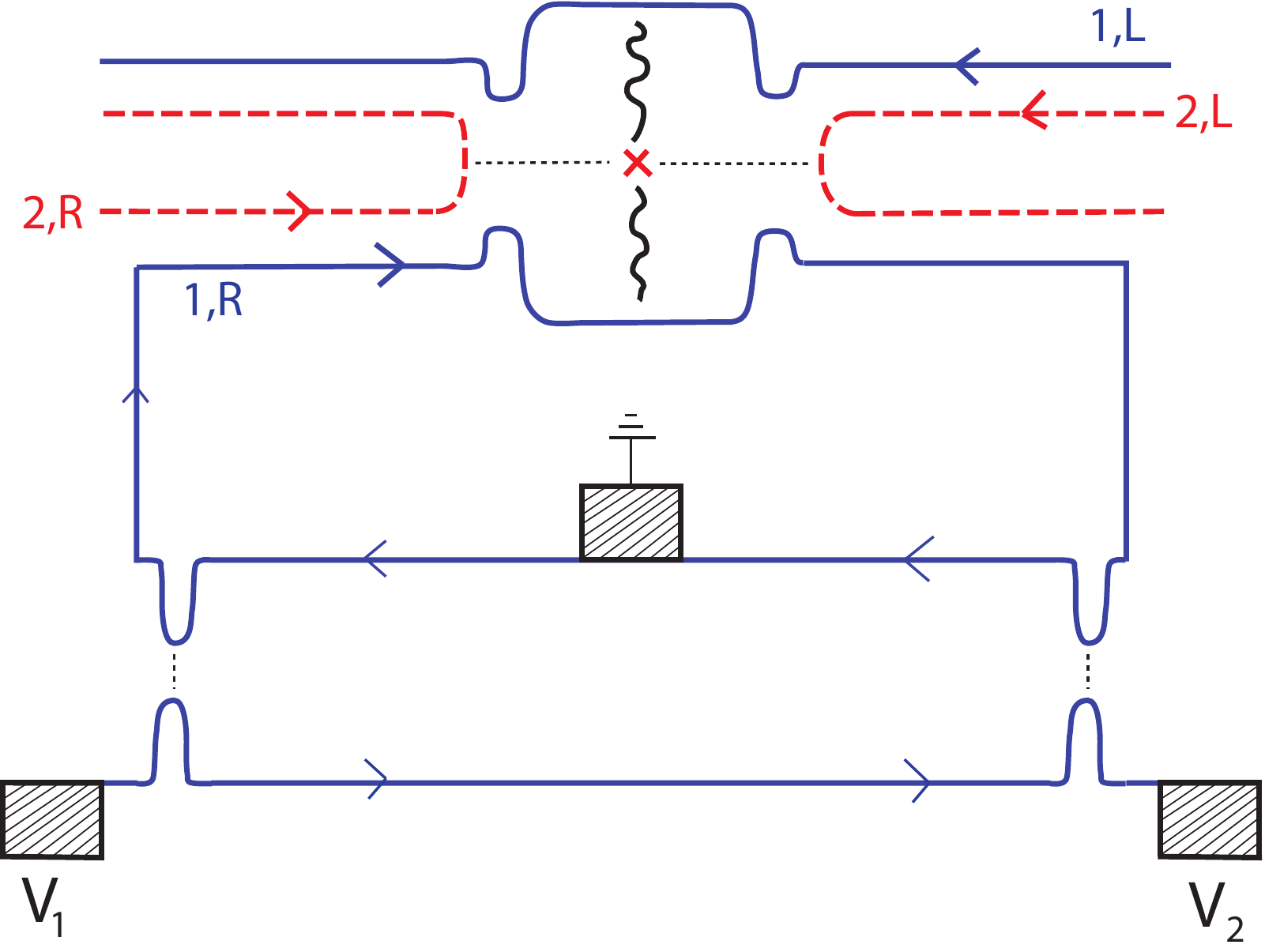}
\caption{Mach-Zehnder interferometer with a quantum dot in its upper arm. The outer edge channel (full line, blue) is fully trasmitted through the dot, the inner edge channel (dashes line, red) has a weak tunnel coupling to the localized state in the dot. If the charge on the localized state is increased by one, the number of electrons on the outer edge is 
accordingly reduced, giving rise to a phase shift $\delta$ . If screening by the two outer edge channels is symmetric, then $\delta = \pi$ and the interference visibility is reduced to zero when the localized level is half filled.}
\label{setup.fig}
\end{figure}

Here we condsider an electronic Mach-Zehnder interferometer (MZI) \cite{heiblum03}, one arm of which is coupled electrostatically to a detector. The interferometer  is defined by the outer edge  channel of a $\nu=2$   quantum Hall setup while the detector consists of   localized electronic state, which is tunnel coupled to   the inner edge, cf. Fig.~\ref{setup.fig}. Our main  findings are (i) Thermal fluctuations of the occupancy of the localized state  lead to dephasing through statistical averaging over shifted interference patterns \cite{heiblum11};  (ii)  In the limit of zero temperature the passage of an electron through the upper arm of the MZI modifies the many-body state of the detector. In similitude to the Anderson orthogonality catastrophy  \cite{Anderson67,Aleiner+97},  the scalar product of the  states before and after this modification has taken  place,
$S_{fi}$,   plays the role of the visibility suppression factor. However, by contrast to the Anderson orthogonality 
catastrophy,  $S_{fi}$ does NOT scale with  the detector's size. Nevertheless, by tuning the gate voltage on the localized impurity,  complete  dephasing can be achieved.   (iii) The strength of the system-detector coupling, hence the degree of dephasing, can be expressed through the Friedel sum rule.The latter may be manipulated by modifying the coupling to the inner edge leads (channels 2L and 2R in Fig.~\ref{setup.fig}). (iv) We briefly discuss the implementation of the Friedel sum rule in the presence of tunnel and/or electrostatic coupling to the localized state whose occupation is varied.  The results (ii) and (iii)   provide a conceptual and technically workable  many-body framework for  dealing with zero temperature dephasing.

{\em Model:} we consider the following Hamiltonian
%
\begin{subequations}
\begin{eqnarray}
& &H_0   =   v \sum_{\alpha=1,2; k} c_{\alpha R,k}^\dagger c_{\alpha R,k} (k - k_F)  - c_{\alpha L,k}^\dagger c_{\alpha L,k} (k + k_F)
\nonumber \\ 
& & + \  \epsilon_0 \;   d^\dagger d\\
& & H_{\rm tunnel}  =  \sum_k \left(\gamma_{2 L,k} c_{2 L,k}^\dagger   + \gamma_{2,R,k} c_{2 R,k}^\dagger 
\right)  d   \ + \ {\rm h.c.}\\
& & H_{\rm imp-edge}  =  \int_{-D/2}^{D/2}\! \! \! \! \! \! \! \! \!  dr \left(\rho_{1 R}(r) V_R(r)  +     \rho_{1 L}(r) V_L(r)     \right) d^\dagger d  \ \ .\nonumber\\ 
\end{eqnarray} 
\end{subequations}
%
Here, $c_{\alpha R/L, k}^\dagger$ creates a left/right moving electron with momentum $k$ on the edge state
$\alpha$, where $\alpha=1$ denotes the outer edge state being part of the Mach-Zehnder interferometer, 
and $\alpha=2$ denotes the inner edge state which is coupled to the localized state inside the qunatum dot. 
 $V_{R/L} (r)$ describes the electrostatic interaction between the dot electron and right/left moving  electrons
 on the outer edge over the interval $[-D/2,D/2]$, and $\rho_{1 R/L}(r)$ are the respective densities. The density operator is normalized such that the equilibrium density $\rho_0$ is subtracted and that the expectation value with respect to $H_0$ vanishes, $\langle \hat{\rho}(r)\rangle_0=0$. $d^\dagger$ creates an electron on the localized state.

{\em Coulomb phase:} when the localized state is occupied with $\langle d^\dagger d\rangle =1$, there is a density 
change 
\begin{equation}
\langle \rho_{R/L}(r) \rangle = - {1 \over 2 \pi v \hbar} \; V_{R/L}(r) \
\end{equation}
on the edges. 
Using standard bosonization, the electron operator on the right moving edge can be expressed as the exponential 
$\psi_{R}(r) = \exp[i \varphi_R(r)]$ with $\rho_R(r) = {1 \over 2 \pi} \partial_r \varphi_R(r)$.  The electron Green function on the right moving lower edge (which is part of the MZ interferometer) can be expressed as

%
\begin{eqnarray}
g_R(x,t) &=& \langle e^{i \varphi_R(x,t) - i \varphi_R(0,0)} \rangle \nonumber \\
& = & e^{2 \pi i \int_{-D/2}^x dy \langle \rho_R(y)\rangle } \langle e^{i \delta \varphi(x,t) - i \delta \varphi(0,0)} \rangle
\end{eqnarray}
where $\delta \varphi(x)  = \varphi(x)  - 2 \pi \int_{-D/2}^x dy \langle \hat{\rho}(y) \rangle$.  If the charge on the impurity is fully screened by the two outer edge states (as is the case for a long-range Coulomb interaction and in the absence of other screening gates besides the edge state), and if the right and left moving electrons contribute equally to the screening, 
then $\int_{-D/2}^{D/2} dy \langle \rho_R(y)\rangle = -1/2$. In order to obtain the correct sign of the phase change as observed when embedding the chiral edge into an MZI \cite{MZFP}, we define the phase shift due to Coulomb coupling as 
%
\begin{equation}
\delta \ = \   {1 \over v \hbar} \int_{-D/2}^{D/2} dy V_R(y) \ \ .
\label{coulomb.eq}
\end{equation}
%
For the case of a symmetric Coulomb coupling of the localized state to two chiral edges as in Fig.~\ref{setup.fig}, 
one finds $\delta = \pi$.

{\em Dephasing by statistical averaging:} the coupling between the localized state and the inner edge 
modes gives a width $\Gamma$ to the localized state. At a finite temperature $T$  larger than the width $\Gamma$ of the QD resonance, 
the occupancy of the localized level  fluctuates thermally with an occupation probability determined by the 
Fermi distribution. In the regime of linear response  $e V_{sd} \ll \Gamma \ll k_B T$, the transmission phase of the 
right moving outer edge channel  depends on the occupancy of the dot. When thermally averaging over the occupancy of the localized state, we find
%
\begin{equation}
\langle e^{i \delta d^\dagger d} \rangle \ = \ {e^{i \delta} \over e^{(\epsilon_0 - \mu)/T} + 1} \ + \ {1 \over 
e^{-(\epsilon_0 - \mu)/T} + 1} \ \ .
\end{equation}
%
In a fully symmetric dot where each of the outer edge modes  does half the screening, we have 
$e^{i \delta}=-1$ and $ \langle e^{i \delta d^\dagger d} \rangle = \tanh((\epsilon_0 - \mu)/T)$. This implies a 
phase lapse of $\pi$ at the degeneracy point $\mu=\epsilon_0$, and a visibility of the interference 
$\nu \propto |\tanh((\epsilon_0 - \mu)/T)|$. The full width at half maximum of the dip in the visibility is 
given by $\approx 2.197\; k_B T$.

{\em Dephasing by entanglement:}  what happens to the visibility of the interference signal when the temperature is much smaller than the width $\Gamma$ of the localized state? More specifically, we will consider the limit where $k_B T \ll \Gamma \ll e V_{sd} \ll \mu - B$, where $B$ denotes the bottom of the band. In this regime, it is useful to discuss the change  an interfering electron on the outer edge makes to its environment, which consists of the localized state coupled to the inner edge. This point of view is equivalent to our discussion of the phase shift $\delta$ of the interfering electron, which is caused by the environment \cite{Stern+90}.  More precisely, we describe the electronic wave function after passage through the interferometer by 
%
\begin{equation}
|u\rangle \otimes |\chi_u\rangle \ +\   e^{i \phi} |d\rangle \otimes |\chi_d\rangle  \ \ .  
\end{equation}
%
Here, $|u\rangle$ and $|d \rangle$ denote the partial waves moving through the upper and lower arm of the MZ interferometer, respectively, $|\chi_u\rangle$ and $|\chi_d\rangle$ denote the respective states of the environment, 
and the Aharonov-Bohm phase $\phi$ is ascribed to the lower arm. As the interference term is the superposition between the partial waves traveling on the upper and lower path, it is reduced by the 
factor $|\langle \chi_u| \chi_d\rangle|$.

Let us start by describing the state $|\chi_d\rangle$, where no interaction has taken place between the interfering electron and the system of inner edge mode and localized state. 
For simplicity, we 
consider a situation where the localized state couples to only one chiral edge mode, $\gamma_{2R,k} \equiv 
\gamma_k$, $\gamma_{2L,k}=0$, but there will only be changes in notation if there is  coupling to two edge modes. We denote an eigenstate of the inner edge plus the localized state    by 
%
\begin{equation}
|\epsilon \rangle = {\cal N} \big( \int_{- D/2}^{D/2}  dx\; \varphi(x) | x\rangle \ + \ A(\epsilon) | d\rangle\big)  \ \ .
\label{epsilon.eq}
\end{equation}
%
Here, $\epsilon$ is the energy of the state, $\varphi(x) $ the wave function along the edge, which suffers a phase shit $\delta_t$ due to the tunnel coupling to the localized state. As the component of the total wave function which describes
the partial wave   traveling along the upper arm is the product of the electronic state $|u\rangle$ and the environment state $|\chi_u\rangle$, the Coulomb phase shift $e^{i \delta}$ can be either assigned to 
the electronic wave function $|u\rangle$, as we did in the last section on thermal dephasing, or to the wave function of the environment $|\chi_u\rangle$ as we will do now. The wave function of the environment is a Slater determinant of  single particle states $|\epsilon\rangle$. However, in these single particle wave functions only the component which has weight on the localized state will be affected by the Coulomb interaction with the intefering electron, such that it gets transformed into 
%
\begin{equation}
|\epsilon_{\delta} \rangle = {\cal N} \big( \int_{- L/2}^{L/2}  dx \varphi(x) | x\rangle \ +\  e^{i \delta}  A(\epsilon) | d\rangle\big)  \ \ .
\label{epsilondelta.eq}
\end{equation}
%
This can be expressed more formally by applying the operator $e^{i \delta d^\dagger d}$ to the state
 $|\epsilon\rangle$. When denoting an expectation value with respect to the ground state Slater determinant 
 of the states $|\epsilon \rangle$ by $\langle ...\rangle$, the overlap between the two states of the environment can be expressed as 
 %
 \begin{equation}
 \langle \chi_u | \chi_d\rangle \ = \ \langle e^{i \delta d^\dagger d} \rangle \ \ .
 \label{fluctuation.eq}
 \end{equation}
 %
Let us discuss the implications of this expression. If the chemical potential is either much below or much above the energy $\epsilon_0$ of the localized level, the occupancy is either 0 or 1, and the number operator 
 $d^\dagger d$ in the exponent can be replaced by its expectation value, thus describing the absence or the presence of the Coulomb phase shift. However, when $|\epsilon_0 - \mu| \approx \Gamma$, the occupancy of the localized level is not a good quantum number. As a consequence, the operator $d^\dagger d$ is not only characterized by its expectation value, but also by all higher cumulants describing quantum flucutations. It is the nonzero value of higher cumulants of $d^\dagger d$ which reduces the expectation value in Eq.~(\ref{fluctuation.eq}). In this sense, dephasing by the quantum dot is intimately related to the existence of quantum 
fluctuations. However, we will explain later that a change in the environment, where all $|\epsilon \rangle $ are 
transformed into $|\epsilon_{\delta}\rangle$, is only possible  if an energy of the order $\Gamma$ is transferred from the interfering electron to the environment, hence the physics we are discussing has nothing to do with dephasing by equilibrium zero point fluctuations.   

In order to calculate the reduction of the interference visibility due to the Coulomb coupling between the interfering electron and the localized state, we need to calculate the matrix of wave function overlaps 
$M_{n,n^\prime} = \langle \epsilon_{n,\delta_C} | \epsilon_{n^\prime}\rangle$. 
After some algebra, we find
%
\begin{equation}
M_{n,n^\prime} \ = \ \delta_{n,n^\prime} \ + \ {\cal N}_{\epsilon_n} A^*(\epsilon_n) {\cal N}_{\epsilon_{n^\prime}}
A(\epsilon_{n^\prime}) \left( e^{i \delta} - 1 \right) \ \ ,
\end{equation}
%
where for local tunneling between the inner edge and the localized state $\gamma_k  \equiv \gamma$
%
\begin{equation}
A(\epsilon) = {\gamma \over \epsilon - \epsilon_0 + i \Gamma} \ \ \ {\rm with } \ \ \ \Gamma = {|\gamma|^2 
\over 2 \hbar v} \ \ .
\end{equation}
%
Following the calculation in \cite{Anderson67}, the overlap $|\langle \chi_u|\chi_d\rangle = {\rm det} |M_{n,n^\prime}|$.  We calculate the determinant by diagonalizing the matrix $M_{n,n^\prime}$. This can be achieved by realizing that it has the structrure $\delta_{n,n^\prime} + \alpha u_n u^*_{n^\prime}$. This type of matrix has one eigenvector with elements $u_n$ and eigenvalue $1 + \alpha \sum_n u^*_n u_n$, and a degenerate  eigenspace which is orthogonal to the vector $\{u_n\}$ and has the  eigenvalue 1. For this reason, the determinant is given by the eigenvalue not equal to one.  Applying this method to the matrix $M_{n,n^\prime}$, we find 
%
\begin{equation}
|\langle \chi_u | \chi_d\rangle | \ = \left| 1 + \left(e^{i \delta} - 1 \right) {\arctan{(\mu - \epsilon_0) 2 \pi \over \Gamma} + {\pi \over 2} \over \pi} \right| \ \ .
\label{reduction.eq}
\end{equation}
%
This reduction of the MZ interference visibility is the central result of our manuscript. 
Interestingly, we note that the overlap $|\langle \chi_u | \chi_d\rangle |$ does not scale with the size of the detector
(i.e.~the length of the inner edge), in sharp contradistinction from factors describing the  Anderson orthogonality
catastrophy. In the case of a symmetric coupling of the outer edge modes to the localized state, and assuming that the charge on the localized state
is fully screened by the outer edge mode, one finds $\delta= \pi$, and as a consequence a complete suppression of the interference visibility for $\mu = \epsilon_0$. On the other hand, for the case of an empty or 
fully occupied localized state, the interference visibility is fully recovered. 

\begin{figure}[h]
\centering\includegraphics[width=0.9\columnwidth]{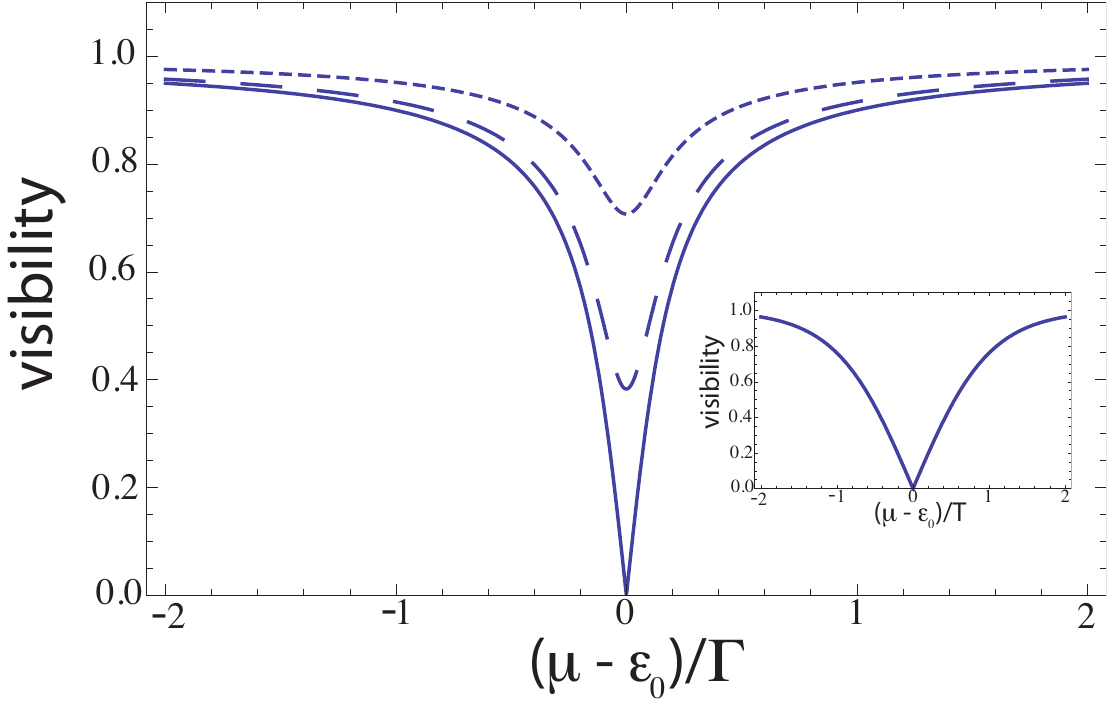}
\caption{Visibility of the interference signal at zero temperature for different values $\delta=\pi$ (full line), $\delta=3 \pi/4$ (dashed line), and $\delta=\pi/2$ (dotted line). Inset: finite temperature dephasing due to a statistical average over different charge states of the localized state. }
\label{toymodel.fig}
\end{figure}

Next, we discuss the condition $e V_{sd} \geqslant \Gamma$ for the applicability of Eq.~(\ref{reduction.eq}). 
We have argued that passage of an interfering electron transforms the Slater determinant of $|\epsilon\rangle$ 
into a Slater determinant of $|\epsilon_{\delta}\rangle$. We now calculate the energy difference between these two many body states. Using the states Eqs.~(\ref{epsilon.eq}), (\ref{epsilondelta.eq}) together with the 
Hamiltonian Eq.~(1), we find for the energy difference 
%
\begin{equation}
\Delta E \ = \ {\Gamma \over \pi} ( 1 - \cos \delta) \ln\left({B^2 + \Gamma^2 \over \mu^2 + \Gamma^2}\right) 
 \ \ .
 \end{equation}
 %
 As the environmental state $|\chi_u\rangle$ is by an amount $\Delta E$ higher in energy than the state
 $|\chi_d\rangle$, this energy difference must be provided by the interfering electron.  As the interfering electron has an energy $e V_{sd}$ above the Fermi level, we arrive at the requirement $e V_{sd}  \geqslant \Gamma  $
 for the interering electron to be able to provide the energy for exciting its environment. 
Due to the fact that an energy transfer is involved in the reduction of MZ interference visibility, we argue that 
this type of dephasing can be described as back-action of the environment onto the interfering electron.  

\begin{figure}[h]
\centering\includegraphics[width=0.9\columnwidth]{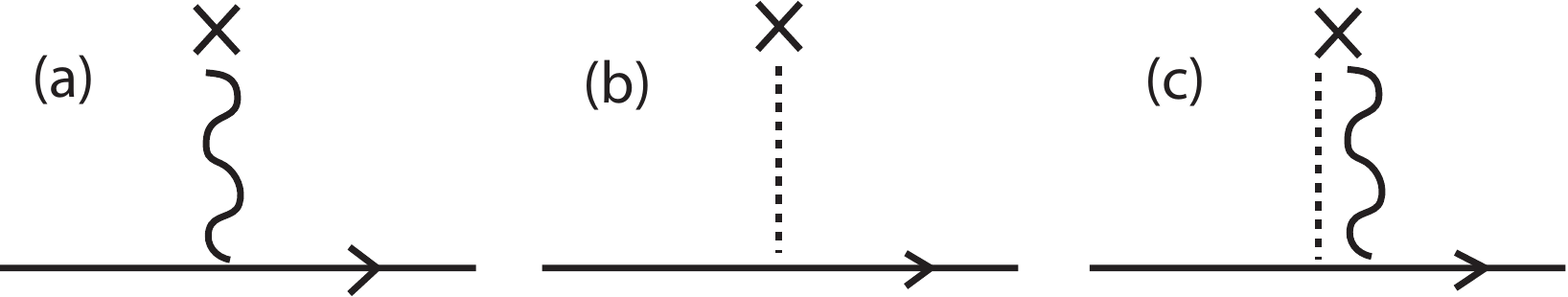}
\caption{Toy model for discussion of Friedel sum rule}
\label{toymodel.fig}
\end{figure}

{\em Friedel sum rule:} our results are intimately related to the Friedel sum rule. 
The Friedel sum rule states that upon the 
change of the charge in a spatially confined region  in a Fermi sea by one,  the sum of 
the scattering phases changes by $2 \pi$, $\Delta N_{\rm confined} = {1 \over 2 \pi i} \Delta {\rm Tr} \ln {\cal S}(\mu)$, 
where ${\cal S}$ is the scattering matrix.  As  an illustration, consider the setup of Fig.~\ref{toymodel.fig}a. In order to make contact with the sign convention of the Friedel sum rule,  we refer here to a chiral edge embedded in a Fabry-Perot interferometer  \cite{MZFP}. When the coupling between  the chiral lead  and the localized impurity state is electrostatic, the latter can be thought of as an external potential which is modified when the impurity state is occupied by an electron. This, in turn, gives rise to a screening cloud ( image charge) on the chiral channel of charge $|e|$; by Friedel sum rule this amounts to a change in the transmission phase through the chiral channel of   $-2\pi$. 

We next consider the chiral lead  and the localized impurity state being tunnel coupled (Fig.~\ref{toymodel.fig}b). Now the impurity state is part of the system. A direct calculation of the scattering phase ( which, for the present setup, is the transmission phase)  yields $   \tan {\delta_{ t} \over 2} = -  {\Gamma \over \epsilon}$ \cite{Mahan}. Filling up the state (i.e., varying $\epsilon$ from $-\infty$ to $+\infty$) results in a total phase change of      $+2\pi$, in agreement with the Friedel sum rule.  Interestingly, when both tunneling and electrostatic coupling are present   (Fig.~\ref{toymodel.fig}c), the two contributions discussed above cancel  each other. This again is in agreement with the Friedel sum rule, as filling up the localized state will induce an opposite sign screening cloud, amounting to a redistribution of charge (with the net total charge unchanged) \cite{Buttiker}.

Now we discuss the more realistic situation in which we have a second chiral. That will correspond to Fig.~\ref{toymodel.fig}a with the provision that the screening cloud of the localized state is now shared with another chiral channel, not shown.  If 
the Coulomb coupling between the localized state and the two chirals is symmetric, the Coulomb phase shift is $-\pi$ for  each of them. We note that the sign of this phase shift is opposite to that in the convention Eq.~(\ref{coulomb.eq}), see \cite{MZFP}. If the coupling to the two chirals is not symmteric, we do not have full dephasing. In the setup Fig.~\ref{setup.fig}, the Coulomb coupling of the localized state is to the outer edge modes, while the localized state is tunnel coupled to the inner edge modes. However, when considering a $4\times 4$ scattering matrix, the sum of changes in scattering phases is again zero as discussed above. While the assumption that all the 
screening of the localized state is done by the outer edge mode is realistic for a strongly pinched inner edge mode, 
it has to be relaxed if the transmission of the inner edge mode through the quantum dot becomes large. In that case, 
the screening  charge is distributed among outer and inner edge modes, the Friedel phase is spread over more channels and reduces the magnitude of the Coulomb phase shift $\delta$. 

In summary, we have discussed dephasing of an MZI interference signal by a detector at equilibrium. At finite 
temperature, dephasing is due to a thermal average over different states of the detector. In the zero temperature 
limit, dephasing is due to quantum fluctuations of the detector in a non-equilibrium interferometer. The strength 
of dephasing is determined by the phase shift caused by Coulomb coupling between detector and interferometer. 
The Friedel sum rule relates this phase shift to the screening charge which an occupied detector state induces on 
the interferometer arm.   

We thank M.~Heiblum for drawing our attention to dephasing by a detector at equilibrium, and we acknowledge
useful discussions with  E.~Weisz. Financial support by the BMBF, The German-Israel foundation (GIF), the Minerva Foundation, and the Israel Science Foundation is acknowledged.

\widetext

\newpage

\subsection*{Supplemental Material: Fabry-Perot interference phase versus Mach-Zender interference phase}
\label{sec:full-transp-stat}

\begin{figure}[tbh]
\centering\includegraphics[width=0.9\columnwidth]{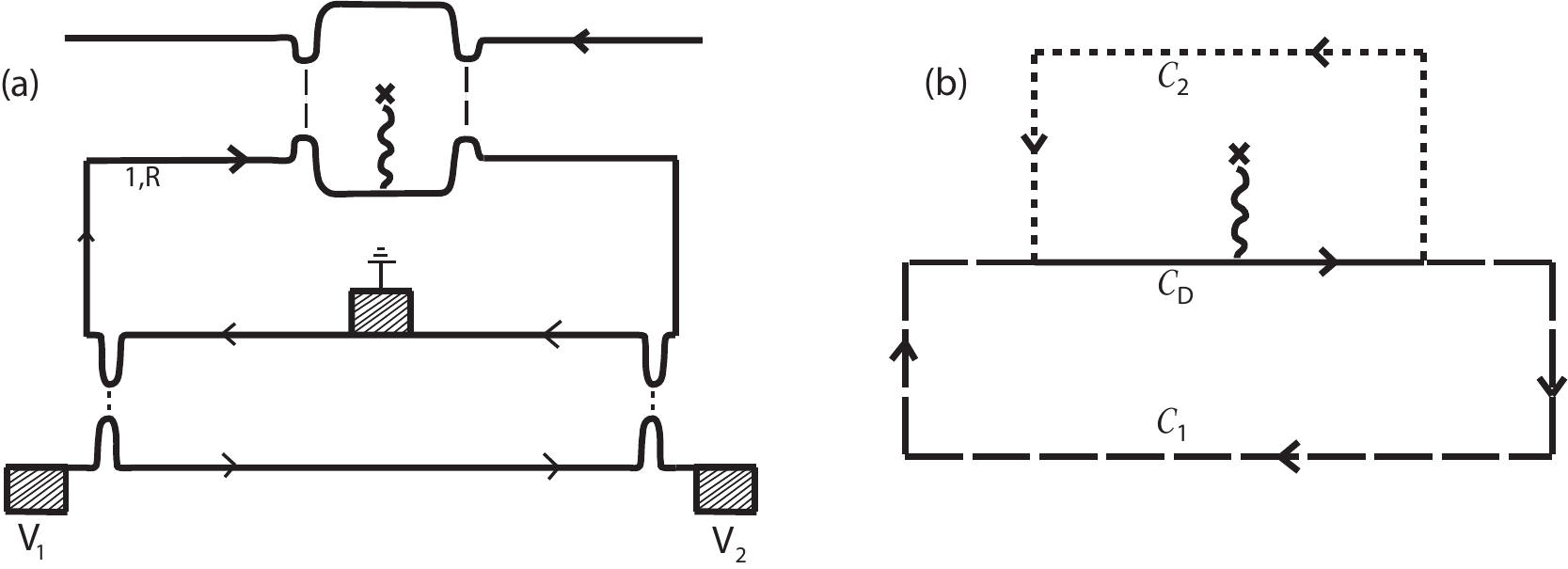}
\caption{Semi-realistic (panel (a)) and idealized (panel (b)) setup for a combined MZI and FPI. Both interferometers share the lower edge of the quantum dot, which is denoted by ${\cal C}_D$ in panel (b). The remaining interference path of the MZI is denoted by ${\cal C}_1$, while the remaining interference path of the 
FPI is denoted by ${\cal C}_2$}
\label{idealized.fig}
\end{figure}

In order to compare the sign of a phase shift due to either Coulomb or tunnel coupling between a Fabry-Perot (FPI) and a Mach-Zehnder  (MZI) interferometer, we use the abstraction of Fig.~\ref{idealized.fig}b. The two interferometers are 
described by the paths along which the partial waves of an interfering electron travel. The path ${\cal C}_{\rm MZI}={\cal C}_1 + {\cal C}_D $ of the MZI has two contributions.  One of them, ${\cal C}_D$, is shared between the MZI and the FPI, whose path is given by ${\cal C}_{\rm FPI}={\cal C}_2 + {\cal C}_D $. In the following, we discuss how a change $\delta$ in the interference phase accrued along ${\cal C}_D$ changes the interference patterns of MZI and FPI, respectively. As edge states are projected onto a single Landau level, 
the interference phase is a pure Aharonov-Bohm phase, and
in the absence of the phase change $\delta$, the MZI and FPI interference phases are given by 
%
\begin{equation}
\varphi_{\rm MZI} = {2 \pi \over \Phi_0} \int_{{\cal C}_1 + {\cal C}_D} \! \! \! \!    d \underline{s} \cdot \underline{A} 
\ < \ 0 \ \ , \ \ \ \varphi_{\rm FPI} = {2 \pi \over \Phi_0} \int_{{\cal C}_2 + {\cal C}_D} \! \! \! \!    d \underline{s} \cdot \underline{A}\ > \ 0 \ \ .
\end{equation}
%
It is important to note that the sign of the two phases is opposite to each other because the sense of revolution around the two inteferometers is opposite. For concreteness, we assume here that the direction of the magnetic field is such that a mathematically positive sense of revolution corresponds to a positive Aharonov-Bohm phase. The phase change $\delta$ is a purely local property and changes only what happens to the interfering electron along the shared segment ${\cal C}_D$. For this reason, it appears as an additive contribution to both interference phases,
%
\begin{equation}
\tilde{\varphi}_{\rm MZI} = {2 \pi \over \Phi_0} \int_{{\cal C}_1 + {\cal C}_D} \! \! \! \!    d \underline{s} \cdot \underline{A} 
\ + \ \delta \ \ , \ \ \ \tilde{\varphi}_{\rm FPI} = {2 \pi \over \Phi_0} \int_{{\cal C}_2 + {\cal C}_D} \! \! \! \!    d \underline{s} \cdot \underline{A}\ + \ \delta  \ \ .
\end{equation}
%
Due to the fact that the original phases $\varphi_{\rm MZI}$ and $\varphi_{\rm FPI}$ differ in sign, the {\em magnitude} of the two phases is affected in opposite ways by the change $\delta$. While the magnitude $|\varphi_{\rm MZI}|$ {\em decreases} by an amount $\delta$, the magnitude  $|\varphi_{\rm FPI}|$ {\em increases} by the same amount. 
Because the area inside the interference loop of an FPI decreases if the interfering edge states screen a negatively charged impurity state, the phase change $\delta_C$ due to Coulomb screening is negative for an FPI. Due to our argument above, this means that the effective phase change due to Coulomb screening is positive for the MZI, as defined in Eq.~(4).  

\end{document}